%% file: paper.tex
\documentclass[aps,prl,unsortedaddress,superscriptaddress,nofootinbib,twocolumn,10pt]{revtex4-2}
\usepackage{natbib}
\usepackage[colorlinks=true,linkcolor=Blue,citecolor=Blue,urlcolor=Blue]{hyperref}
\usepackage[utf8]{inputenc}
\usepackage[dvipsnames,x11names]{xcolor}
\usepackage{amsfonts,amsmath,amssymb}
\usepackage{graphicx}
\usepackage{bm}
\usepackage[normalem]{ulem}
\usepackage{xspace}
\usepackage{ifthen}
\usepackage{mathtools}
\usepackage{newtxtext}
\usepackage{newtxmath}
\usepackage[T1]{fontenc}
\usepackage{orcidlink}

\input{macros}

\newcommand{\EndMatter}[1][End Matter]{%
  \par
  \onecolumngrid
  \vspace{1.2\baselineskip}
  \begin{center}
    \rule{0.45\textwidth}{0.4pt}\\[0.9\baselineskip]
    {\bfseries\MakeUppercase{#1}}
  \end{center}
  \vspace{0.4\baselineskip}
  \twocolumngrid
}
\makeatletter
\renewcommand\onecolumngrid{%
\do@columngrid{one}{\@ne}%
\def\set@footnotewidth{\onecolumngrid}%
\def\footnoterule{\kern-6pt\hrule width 1.5in\kern6pt}%
}
\renewcommand\twocolumngrid{%
        \def\footnoterule{%
        \dimen@\skip\footins\divide\dimen@\thr@@
        \kern-\dimen@\hrule width.5in\kern\dimen@}
        \do@columngrid{mlt}{\tw@}
}%
\AtBeginDocument{%
  \setlength{\@fptop}{0pt}\setlength{\@fpsep}{12pt plus 1pt}\setlength{\@fpbot}{0pt plus 1fil}%
  \setlength{\@dblfptop}{0pt}\setlength{\@dblfpsep}{12pt plus 1pt}\setlength{\@dblfpbot}{0pt plus 1fil}}
\makeatother

\usepackage[compact]{titlesec}
\titleformat{\section}[runin]{\itshape}{}{0pt}{}[.---]
\titlespacing*{\section}{0pt}{*1}{0pt}

\begin{document}

\title{\boldmath \papertitle}

\newcommand{\ific}{\affiliation{\small%
Instituto de F\'isica Corpuscular, Centro Mixto Universidad de Valencia-CSIC, \\
Institutos de Investigaci\'on de Paterna, Apartado 22085, E-46071, Valencia, Spain}}

\newcommand{\teoUV}{\affiliation{\small%
Departamento de F\'\i sica Te\'orica and IFIC, Centro Mixto Universidad de Valencia-CSIC,
Institutos de Investigaci\'on de Paterna, Aptdo. 22085, E-46071 Valencia, Spain}}

\author{Miguel Albaladejo\orcidlink{0000-0001-7340-9235}}
\email{Miguel.Albaladejo@ific.uv.es}
\ific

\author{Pablo Encarnación\orcidlink{0009-0005-0749-3885}}
\email{Pablo.Encarnacion@ific.uv.es}
\teoUV

\author{Albert Feijoo\orcidlink{0000-0002-8580-802X}}
\email{Eduardo.Feijoo@ific.uv.es}
\teoUV

\author{Juan Nieves\orcidlink{0000-0002-2518-4606}}
\email{Juan.M.Nieves@ific.uv.es}
\ific

\begin{abstract}
The LHCb Collaboration has observed a narrow state in the $\Bsbar\piz$ spectrum, about $74\MeV$ below the
$\Bm\Kp$ threshold, which is the natural candidate for the bottom partner of the $\Dsstar$. Such a state had been
predicted as a $\Bbar K$ bound state by unitarized heavy-meson chiral perturbation theory with lattice-QCD input. 
We analyze the LHCb spectrum with these coupled-channel amplitudes, including isospin breaking through the physical meson masses, $\piz$--$\eta$ mixing, and $m_u\neq m_d$ in the chiral amplitudes, thereby allowing the $\Bsstar$ to decay through its only open strong channel, $\Bsbar\piz$. With the parameters fixed by
lattice QCD and no experimental input from the LHCb measurement, the predicted state lies within
$1.1$--$1.7$ standard deviations of the observed peak. Adjusting a single parameter, the
spectrum is described as well as with the zero-width limit of the LHCb analysis. The new state thus emerges naturally from these
amplitudes, with a large $\Bbar K$ component and an isospin-violating width of tens of keV, and its $J^P=1^+$
heavy-quark spin partner is predicted with a precision comparable to that of the measured mass.
\end{abstract}

\maketitle

\section{Introduction}
The discovery of the $\Dsstar$ and $\Dsone$ mesons in the $D_s^{(*)+}\piz$ spectra~\cite{BaBar:2003oey,CLEO:2003ggt}
opened one of the longest debates in hadron spectroscopy. Their masses lie about $150\MeV$ below the quark-model
expectations for the $0^+$ and $1^+$ $c\bar{s}$ states~\cite{Godfrey:1985xj}, just below the $DK$ and $D^*K$
thresholds, and they are very narrow, because their only open strong decays, into $D_s^{(*)}\piz$, violate
isospin~\cite{Cho:1994zu}. They have been interpreted as the chiral partners of the ground-state
$D_s^{(*)}$ mesons~\cite{Bardeen:2003kt,Nowak:2003ra}, as compact tetraquarks~\cite{Cheng:2003kg}, and as
hadronic molecules generated by the $D^{(*)}K$ interaction~\cite{Barnes:2003dj,Kolomeitsev:2003ac,Guo:2006fu,Guo:2006rp,Gamermann:2006nm}
(see Ref.~\cite{Guo:2017jvc} for a review). The molecular picture has received support from lattice QCD
(LQCD): the finite-volume $DK$ and $D^*K$ levels~\cite{Mohler:2013rwa,Lang:2014yfa,Bali:2017pdv}, analyzed with
unitarized chiral amplitudes, give a sizable $D^{(*)}K$ component~\cite{MartinezTorres:2014kpc,Albaladejo:2018mhb,Gil-Dominguez:2023puj,Wang:2025tgl}, as do the $DK$ invariant-mass
distributions measured in $B$ decays~\cite{Albaladejo:2016hae}. Moreover, the
scattering lengths of light pseudoscalars off charmed mesons computed in LQCD~\cite{Liu:2012zya} fix the
next-to-leading-order (NLO) low-energy constants (LECs) of heavy-meson chiral perturbation theory (HMChPT), whose
unitarized amplitudes generate the $\Dsstar$ as a $DK$ bound state and describe the positive-parity charmed
mesons~\cite{Liu:2012zya,Albaladejo:2016lbb,Du:2017zvv}. The isospin-violating widths have also been computed in the
molecular picture~\cite{Lutz:2007sk,Guo:2008gp,Liu:2012zya,Fu:2021wde}. The $\Dsstar$ has thus become a paradigm of how chiral dynamics near a threshold can reshape the hadron spectrum.

Heavy-quark flavor symmetry (HQFS) relates the charm and bottom sectors, and heavy-quark spin symmetry (HQSS)~\cite{Isgur:1989vq,Isgur:1990yhj,Georgi:1990um,Isgur:1991wq,Manohar:2000dt} the
$0^+$ and $1^+$ states; the same approaches predicted $0^+$ and $1^+$ $\Bbar{}_s$ states below the $\Bbar{}^{(*)}K$
thresholds~\cite{Bardeen:2003kt,Nowak:2003ra,Kolomeitsev:2003ac,vanBeveren:2003af,Mehen:2005hc,Colangelo:2005gb,Guo:2006fu,Guo:2006rp,WooLee:2006kdh,Badalian:2007yr,Guo:2007up,Cleven:2010aw,Guo:2011dd,Colangelo:2012xi,Dmitrasinovic:2012zz,Altenbuchinger:2013vwa,Torres-Rincon:2014ffa,Cheng:2014bca,Cleven:2014oka,Albaladejo:2016lbb,Ortega:2016pgg,Albaladejo:2016ztm,Cheng:2017oqh,Du:2017zvv,Sun:2018zqs,Zhou:2020moj,Alhakami:2020vil,Guo:2021rjv,Fu:2021wde,Gandhi:2022nnk,Yang:2022vdb,Kim:2023htt,Ni:2023lvx,Zhang:2024usz,Hao:2025vmw}
(see Ref.~\cite{Chen:2016spr} for a review). In particular, the unitarized NLO HMChPT amplitudes with the LECs
fixed by the charm-sector LQCD data predicted the $\Bsstar$ at $5724^{+17}_{-24}\MeV$~\cite{Albaladejo:2016lbb}
[$5720^{+16}_{-23}\MeV$ in Ref.~\cite{Du:2017zvv}], and a framework combining the leading-order (LO) $\Bbar{}^{(*)}K$
interaction with the bare $b\bar{s}$ states of a constituent quark model, fitted to the $\Bbar{}^{(*)}K$ LQCD levels of
Ref.~\cite{Lang:2015hza}, gave $5711\pm6$ and $5707\pm6\MeV$ for two sets of bare
masses, (a) and (b), respectively~\cite{Albaladejo:2016ztm}. A recent analysis of the $\Bbar{}^{(*)}K^{(*)}$ coupled channels in the
hidden-gauge approach placed these states considerably higher, at $5760$ ($0^+$) and $5802\MeV$
($1^+$)~\cite{Sanchez-Illana:2026cyv}, above both these predictions and the mass later measured by LHCb. These states have also been studied with QCD sum
rules~\cite{Wang:2007tu,Wang:2008tm,Bracco:2010bf,Wang:2015mxa,Zhou:2025yjb} and in LQCD~\cite{Green:2003zza,Koponen:2007nr,Burch:2008qx,Jansen:2008si,Michael:2010aa,Gregory:2010gm,Lang:2015hza,Wurtz:2015mqa,Hudspith:2023loy,Gayer:2024akw,Guyton:2025pma},
\textit{e.\,g.}\,Ref.~\cite{Lang:2015hza} finds a $0^+$ state at $5.711\pm0.023\GeV$, below the $\Bbar K$ threshold. Their
isospin-violating widths were also estimated~\cite{Faessler:2008vc,Feng:2011zzb,Fajfer:2016xkk,Fu:2021wde}, and
processes to observe them were proposed~\cite{Fu:2021wde,Yuan:2025qgv,Fu:2026ouv}. A narrow $0^+$ state below the
$\Bbar K$ threshold, decaying only into $\Bsbar\piz$, was thus a well-defined prediction awaiting experimental test.

Recently, the LHCb Collaboration has observed a narrow structure in the $\Bsbar\piz$ mass spectrum, with
$M_{\rm LHCb}=5698.9\pm1.5\pm0.6\MeV$, a natural width compatible with zero ($\Gamma<9.8\MeV$ at 90\% CL) and
$343^{+44}_{-41}$ signal candidates~\cite{LHCb:2026knl}. Interpreted as the $0^+$ state, it lies far below the
quark-model expectations~\cite{Ebert:1997nk,Lahde:1999ih,DiPierro:2001dwf,Lakhina:2006fy,Vijande:2007ke,Ebert:2009ua,Sun:2014wea,Godfrey:2016nwn,Lu:2016bbk,Asghar:2018tha,li:2021hss,Jakhad:2025ejc}, but only $8\MeV$ below the
prediction $5707\pm6\MeV$ of Ref.~\cite{Albaladejo:2016ztm}. It also lies about $74\MeV$ below the $\Bm\Kp$ threshold,
so that $\Bsbar\piz$ is its only open strong decay channel, as $D_s^+\piz$ for the $\Dsstar$.
(LHCb also considers the $1^+$ state decaying into $\olsi{B}{}_s^{*0}\piz$ with an unreconstructed photon, for which the
mass would be $5748.3\MeV$~\cite{LHCb:2026knl}.) Since its announcement, the state has been discussed as the chiral
partner of the $\Bsbar$~\cite{Nowak:2026zhc}, and two LQCD calculations extrapolated to the continuum limit and to the
physical point have found the $\Bsstar$ bound by $65.9(6.0)(3.0)\MeV$~\cite{Hudspith:2026vxt} and
$69(13)(4)\MeV$~\cite{Guyton:2026mja} with respect to the $\Bbar K$ threshold, the latter concluding that the
positive-parity states are predominantly molecular.

The bottom-strange sector is thus a test of the chiral dynamics of heavy-light mesons, which is not settled in the charm sector: the ALICE Collaboration has measured the $D\pi$ and $D^*\pi$ femtoscopic correlation functions and found them compatible with a negligible strong interaction~\cite{ALICE:2024bhk}, in tension with theoretical calculations incorporating the unitarized NLO HMChPT amplitudes~\cite{Albaladejo:2023pzq,Torres-Rincon:2023qll}, that generate the $\Dsstar$ as a $DK$ bound state and a two-pole pattern for the $D^*_{0}(2300)$ (see also Ref.\,\cite{Encarnacion:2026toappear}). The new beauty-strange state discovered by LHCb offers an independent test of the same amplitudes, in a sector where HQSS is better realized.

In this Letter, we confront the LHCb spectrum with the two unitarized chiral approaches discussed above: NLO HMChPT and the LO interaction with an explicit bare $b\bar{s}$ state, both constrained by LQCD and predicting the $\Bsstar$ as a pole. The observed signal requires isospin breaking, which we incorporate through the physical meson masses, $\piz$--$\eta$ mixing, and $m_u\neq m_d$ in the chiral amplitudes. We compute the invariant-mass distribution from the final-state interaction of $\Bbar K$ pairs, fold it with the detector resolution, and fit the LHCb spectrum, first keeping the unitarized chiral amplitudes fixed and then varying one of their parameters. A good description of the spectrum does not, by itself, establish the molecular nature of the state. We instead test whether the amplitudes that predicted it also describe the data consistently. We find that they do, with only a minimal adjustment of LQCD-constrained parameters, yielding the properties of the observed state and predicting those of its spin partner, anchored by the measured mass.

\section{Amplitudes}
We work in the charged-particle basis, with the $J^P=0^+$ channels $\Bsbar\piz$, $\Bm\Kp$, $\Bz\Kz$ and
$\Bsbar\eta$, and the physical (PDG) masses~\cite{PDG2026}. Isospin is broken by the mass differences, which split the two $\Bbar K$ thresholds and loops,
and by $\piz$--$\eta$ mixing, with the LO angle $\varepsilon=0.0121$ obtained from the FLAG quark-mass
ratios~\cite{FLAG:2024}. Electromagnetic corrections are not included. The $S$-wave amplitudes follow from the
on-shell Bethe--Salpeter equation~\cite{Oller:1998zr,Nieves:1999bx},
\begin{equation}
T(s)=\big[1-V(s)\,G(s)\big]^{-1}V(s),
\label{eq:BSE}
\end{equation}
with $s$ the square of the total energy in the center-of-mass frame and $G$ the diagonal matrix of two-meson loop
functions. We consider two models, both constrained by LQCD, whose
potentials are particular cases of
\begin{align}
f^2V_{ij}={}&\tfrac14C_{ij}\,(s-u_{ij})
+\frac{2c^2F_iF_j\,\mbare\sqrt{m_{H,i}m_{H,j}}\,E_{L,i}E_{L,j}}{s-\mbare^2}\nonumber\\
&-4C^0_{ij}h_0+2C^1_{ij}h_1-2C^{24}_{ij}H_{24}+2C^{35}_{ij}H_{35},
\label{eq:V}
\end{align}
with $f=92.21\MeV$, $m_{H(L),i}$ the heavy (light) meson mass of channel $i$ and $E_{L,i}$ the c.m.\ energy of its
light meson. The first term is the LO [Weinberg--Tomozawa (WT)] HMChPT potential~\cite{Wise:1992hn,Yan:1992gz},
common to both models; the flavor coefficients $C_{ij}$, $C^{0,1,24,35}_{ij}$ and $F_i$ are derived from the chiral
Lagrangians in the charged basis~\cite{supp}.

\emph{Model \modelB}~\cite{Albaladejo:2016ztm} contains the first line of Eq.~\eqref{eq:V}: the WT term and the
$s$-channel exchange of a bare $b\bar{s}$ state of mass $\mbare$. In Ref.\,\cite{Albaladejo:2016ztm} two choices for this mass are considered, $\mbare=5851\MeV$ (set \setA) and $\mbare=5801\MeV$ (set \setB), taken from constituent quark models. For $\varepsilon\neq0$ the bare state also couples directly to $\Bsbar\piz$. The LEC $c$ and the Gaussian regulator $\Lambda$ of the loops were fitted to the $0^+$ and $1^+$ $\Bbar{}^{(*)}K$ LQCD levels of Ref.~\cite{Lang:2015hza}. The $\Bsbar\eta$ channel is not included in this model, so that the isospin limit is exactly the model of Ref.~\cite{Albaladejo:2016ztm}.

\emph{Model \modelC}~\cite{Liu:2012zya,Albaladejo:2016lbb} contains the WT term and the NLO terms of the second line. The four channels are included, with $h_{0,1}$ multiplying the light-quark mass insertions and $H_{24}$, $H_{35}$ containing the two-derivative LECs $h_{2\text{--}5}$~\cite{Liu:2012zya}. The loops are dimensionally regularized with a common subtraction constant $a_c$. The LECs, including $a_c$, were determined from the LQCD scattering lengths of Ref.~\cite{Liu:2012zya} in the charm sector and translated to the bottom sector by HQFS~\cite{Albaladejo:2016lbb}. Besides the mass differences and $\piz$--$\eta$ mixing, the $h_{0,1}$ terms contain a direct isospin-breaking part proportional to $m_u-m_d$, also fixed by the same FLAG input. Thus, isospin breaking introduces no additional free parameters in either model.

In both models the $\Bsstar$ is a pole below the $\Bbar K$ thresholds, $\sqrt{s_p}=M_p-i\Gamma/2$, on the sheet
reached by crossing the $\Bsbar\piz$ cut. Its width is the isospin-violating
$\Gamma(\Bsstar\to\Bsbar\piz)=|g_{\Bsbar\piz}|^2p_{\piz}/(8\pi M_p^2)$, of the order of tens of keV, far below the experimental resolution. The uncertainties of the LQCD-constrained parameters
are propagated by a bootstrap of the lattice data (see the End Matter for technical details). With fixed parameters,
the poles of both models lie at the positions quoted in the Introduction, up to small isospin-breaking corrections. Model~\modelB also contains a broad resonance at $6.2$--$6.3\GeV$~\cite{Albaladejo:2016ztm},
which leaves no visible trace in the $\Bsbar\piz$ spectrum~\cite{supp}.

\section{Production and spectrum}
The $\Bsbar\piz$ pairs of the LHCb sample are produced in $pp$ collisions. We assume a spinless source producing meson
pairs in $S$ wave at short distances, followed by the final-state interaction:
\begin{equation}
\mathcal A(\sqrt s)=\sum_j\alpha_j\big[\delta_{j1}+G_j(s)\,T_{j1}(s)\big],
\label{eq:prod}
\end{equation}
with $j=1$ the $\Bsbar\piz$ channel and $\sqrt s$ the invariant mass. We take $\Bbar K$ produced in $I=0$, that is, $\alpha_{\Bm\Kp}=\alpha_{\Bz\Kz}$. Furthermore, in the \emph{fixed-amplitude} fits, allowing direct $\Bsbar\piz$ or $\Bsbar\eta$ production would turn the predicted peak into a smooth continuum absorbed by the background, while in the \emph{free-amplitude} fits, when the pole position varies, these terms have a negligible impact on the fit~\cite{supp}. We therefore take $\alpha_{\Bsbar\piz}=\alpha_{\Bsbar\eta}=0$.

The distribution is $\dd N/\dd\sqrt s\propto(p_1/\sqrt s)|\mathcal A|^2$, with $p_1$ the $\Bsbar\piz$ momentum. Near the pole, at $\sqrt s\simeq M_p$, the number of events in the peak is
\begin{equation}
N_{\rm peak}\propto\frac{4\pi^2}{M_p}\Big|\sum_j\alpha_jG_j(M_p^2)\,g_j\Big|^2 ,
\label{eq:Npeak}
\end{equation}
independent of the small isospin-violating coupling $g_{\Bsbar\piz}$: below the $\Bbar K$ thresholds the state decays
into $\Bsbar\piz$ with a branching ratio of one, and the yield is driven by the production of $\Bbar K$ pairs, which
couple strongly to it. This is what makes the state visible: the tiny isospin-violating coupling fixes its width, but
not the number of events. The signal is a narrow peak at $M_p$ on top of a continuum smaller by several orders of
magnitude, so that the spectrum constrains essentially the pole mass and the yield $\yield$, which we treat as a free
normalization.

LHCb describes the signal as a relativistic $S$-wave Breit--Wigner function convolved with mode-dependent detector
response functions, and finds the natural width consistent with zero, the likelihood being maximal at
$\Gamma=0$~\cite{LHCb:2026knl}. Since the width is negligible compared with the resolution, the signal component of their fit to the spectrum summed over the three decay modes provides, up to a normalization, the effective resolution function of this spectrum,
\begin{equation}
R(E-M_{\rm LHCb})\propto S_{\rm LHCb}(E),
\label{eq:R}
\end{equation}
with $M_{\rm LHCb}$ the mass measured by LHCb. We use this $R$ to fold our theoretical distributions. Replacing it by a Gaussian
function of the same full width at half maximum, or by a double-sided Crystal Ball function fitted to it, changes very
little our results~\cite{supp}.

The expected number of events in each of the $10\MeV$ bins is
$\mu_b=S_b+B_b$, with $S_b$ the folded theoretical signal normalized to $\yield$ events~\cite{supp} and $B_b$ the background,
taken as in the LHCb analysis~\cite{LHCb:2026knl}: a sigmoid function modulated by a first-order polynomial with four
free parameters. We minimize the binned
Poisson deviance~\cite{BakerCousins1984}
\begin{equation}
D=2\sum_b\big[\mu_b-n_b+n_b\ln(n_b/\mu_b)\big],
\label{eq:deviance}
\end{equation}
with $n_b$ the LHCb candidates.
As a reference, we fit the LHCb spectrum with a zero-width peak of free position $M_0$ and yield $\yield$, folded with
the resolution function $R$, on top of the same background with all its parameters free. This fit is equivalent to the $\Gamma\to0$
limit of the LHCb analysis and yields $M_0=5699.6^{+1.6}_{-1.7}\MeV$ and $\yield=343^{+45}_{-39}$, in agreement with the LHCb
values, $5698.9\pm1.5\MeV$ (statistical uncertainty) and $343^{+44}_{-41}$. The minimum of the deviance of this fit is
the benchmark against which our fits are compared.

We perform two types of fits. In the \emph{fixed-amplitude} fits (\fitX), the parameters of the amplitudes are those
determined from LQCD, the yield $\yield$ is fixed to the LHCb value and only the background is fitted: they test the
predictions as they stand. In the \emph{free-amplitude} fits (\fitY), $\yield$ and the background are fitted together with one parameter of the amplitude: the cutoff $\Lambda$ (Model~\modelB) or the subtraction constant $a_b$ (Model~\modelC), with the latter determined independently of $a_c$, the charm-sector value. The other amplitude parameters, the coupling $c$ (Model~\modelB) and the NLO LECs (Model~\modelC), are taken from the bootstrap samples obtained in Refs.~\cite{Albaladejo:2016ztm} and \cite{Liu:2012zya}, respectively, by fitting only LQCD results. In this way, the theoretical-model uncertainties are also propagated to the event distribution.

\section{Results}
\input{fig_specX}

Figure~\ref{fig:specX} shows the \emph{fixed-amplitude} fits.
When the parameters of the amplitude are not fitted, the pole of Model~\modelC lies above the mass obtained by LHCb,
although close to being compatible with it within the uncertainty of our prediction~\cite{Albaladejo:2016lbb}. This uncertainty, propagated to the event distribution, produces the large error bands of the histogram, and the description of the data, within errors, is not far from the experiment. Model~\modelB, which is even
closer to the data, has narrower error bands because the uncertainty of the predicted pole is
smaller. Quantitatively, the LHCb mass lies at $1.1\,\sigma_\text{\modelC}$ from the prediction of Model~\modelC and
at $1.7\,\sigma_\text{\modelB}$ and $1.3\,\sigma_\text{\modelB}$ from those of Model~\modelB, sets \setA and \setB, respectively (see Table~\ref{tab:results}). These predictions predate the observation, with parameters fixed by LQCD simulations at unphysical light-quark masses and, for Model~\modelC, constrained by LQCD only in the charm sector, without including systematic uncertainties. Since they use no experimental information from the bottom-strange sector, their agreement with the measurement at the one-to-two-standard-deviation level provides a nontrivial test of these approaches.

\input{fig_specY}

Figure~\ref{fig:specY} shows the \emph{free-amplitude} fits. Adjusting a single parameter, both models describe the spectrum
as well as the reference LHCb-like zero-width peak ($\chi^2/\mathrm{dof}=1.22$ for $84$ degrees of freedom,
interpreting the deviance as a $\chi^2$), and the pole follows the data, at
$M_p=5699.6^{+1.6}_{-1.7}\MeV$ in both models (Model~\modelB, set \setB, and Model~\modelC), the central values differing by
around $0.01\MeV$. The required changes are moderate: the subtraction constant of Model~\modelC shifts by $-0.051^{+0.044}_{-0.042}$ from its LQCD value ($1.11\,\sigma$), while $\Lambda$ of Model~\modelB shifts from its LQCD value, $654\pm33\MeV$, to $708\pm27\MeV$, corresponding to $1.28\,\sigma$ of its LQCD distribution. Refitting the LHCb spectrum and LQCD data simultaneously (variants \modelBp and \modelCp, End Matter) yields the same pole positions, at only a small cost in the lattice-data description: $\chi^2_{\rm LQCD}$ increases by $1.3$ for the six finite-volume levels (\modelBp, set \setB) and by $1.1$ for the 15 scattering lengths (\modelCp). Thus, the same amplitudes simultaneously describe the LHCb spectrum and lattice results, fixing the properties of the state listed in Table~\ref{tab:results}.

\input{tab_results}

The table gives the $\Bbar K$ compositeness, $X_{\Bbar K}=X_{\Bm\Kp}+X_{\Bz\Kz}$, with
$X_j=-g_j^2\,\dd G_j/\dd s$ at the pole~\cite{Weinberg:1965zz,Gamermann:2009uq,Aceti:2014ala,Albaladejo:2022sux} (End Matter).
In Model~\modelB it is about one half, as found in Ref.~\cite{Albaladejo:2016ztm}, the rest being the bare
$b\bar{s}$ component. In Model~\modelC, without a bare state, it is larger, around $60\%$.

The width, $\Gamma=82.5^{+12.3}_{-10.3}\keV$ (Model~\modelC) and
$34.0^{+0.3}_{-0.2}\keV$ (Model~\modelB), is two orders of magnitude below the LHCb upper limit; it arises from the
constructive interference of $\piz$--$\eta$ mixing and the $\Bbar K$ mass differences, and, in Model~\modelC, from the
direct $m_u\neq m_d$ terms. It should be regarded only as an estimate (this is similar to the $T_{cc}^+$ case, see
Refs.~\cite{LHCb:2021vvq,LHCb:2021auc,Albaladejo:2021vln,Du:2021zzh}). Unlike the mass and the yield [Eq.~\eqref{eq:Npeak}], $\Gamma$ is proportional to $|g_{\Bsbar\piz}|^2$, which is sensitive to the details of isospin breaking discussed above. It is therefore sensitive to parameters essentially fixed by the amplitudes rather than fitted. The quoted intervals include only parametric uncertainties, while the factor of about two between the two models provides an estimate of the model dependence. Moreover, the radiative decay $\Bsstar\to\olsi{B}{}_s^{*}\gamma$, not included here, contributes to the total width~\cite{Faessler:2008vc,Fu:2021wde} (see Ref.~\cite{Aliev:2026zyc} for the radiative decays of the $1^+$ partner). Our values are of the same order as previous estimates~\cite{Fajfer:2016xkk,Fu:2021wde}.

HQSS relates the $0^+$ state to a $1^+$ partner, bound in $\Bbar{}^*K$ and generated by the same
interaction. In the \emph{free-amplitude} fits its mass is correlated with the observed one: $M(1^+)=5748.8^{+1.6}_{-1.7}\MeV$
(Model~\modelB) and $5751.7^{+1.9}_{-2.2}\MeV$ (Model~\modelC). These values are close to $5748.3\pm1.5\pm0.6\MeV$, the mass
that LHCb obtains if the peak is instead the $1^+$ state, decaying into $\olsi{B}{}_s^{*0}\piz$ with the photon of
$\olsi{B}{}_s^{*0}\to\Bsbar\gamma$ unreconstructed~\cite{LHCb:2026knl}. The reason is that the predicted $1^+$--$0^+$
splitting, $49.26\pm0.01\MeV$ and $52.12^{+1.06}_{-1.28}\MeV$, obtained in the \emph{free-amplitude} fits with models \modelB{}\,\setB and \modelC, respectively, is close to the
$\olsi{B}{}_s^{*}$--$\olsi{B}{}_s$ mass difference that separates the two LHCb values, $49.4\MeV$. This is expected from
HQSS~\cite{Du:2017zvv}, and depends little on the LQCD-constrained parameters. (The \emph{free-amplitude} fit with Model \modelB{}\,\setA gives a smaller splitting, $41.45^{+0.04}_{-0.03}\MeV$, because its bare $b\bar{s}$ $0^+$ and $1^+$ masses, $5851$ and $5883\MeV$, are closer than those of set \setB, $5801$ and $5858\MeV$.\footnote{The bare masses of set \setB incorporate corrections to the one-gluon exchange potential affecting mainly the $0^+$ sector~\cite{Gupta:1994mw,Lakhina:2006fy}.})
Bearing this in mind, two consequences follow. First, interpreting the peak as the $1^+$ state would yield a $0^+$ state within a few MeV of the one obtained here. The $(0^+,1^+)$ doublet predicted by our amplitudes is nearly unchanged under either assignment of the observed peak, which the $\Bsbar\piz$ mass alone cannot distinguish. Second, the $1^+$ state, through its $\olsi{B}{}_s^{*0}\piz$ decay with the photon unreconstructed, would produce a peak at practically the same position as the $0^+$ state. The observed structure could therefore contain both states, as considered by LHCb~\cite{LHCb:2026knl}, a possibility that arises naturally in our framework; their relative yields are beyond the scope of this work. The similar binding energies of the $0^+$ and $1^+$ states agree with recent LQCD calculations, which find them bound by $65.9(6.0)(3.0)$ and $60.6(6.6)(3.0)(1.0)\MeV$~\cite{Hudspith:2026vxt}, or by $69(13)(4)$ and $77(10)(5)\MeV$~\cite{Guyton:2026mja}, below the $\Bbar K$ and $\Bbar{}^*K$ thresholds, respectively. Our amplitudes thus provide a consistent picture of the whole $(0^+,1^+)$ doublet, anchored to the measured mass.

\section{Summary}
We have analyzed the LHCb $\Bsbar\piz$ spectrum using two coupled-channel HMChPT approaches: an LO formulation with an explicit bare $b\bar{s}$ state and an NLO one. Both models have parameters constrained by LQCD and incorporate isospin-breaking effects. In either approach, the observed state appears as a pole below the $\Bbar K$ thresholds, with its yield set by $\Bbar K$-pair production and its isospin-violating width of order tens of keV. The amplitudes fixed before the observation place the pole within $1.1$--$1.7$ standard deviations of the measured mass, while a moderate adjustment of a single parameter describes the spectrum as well as the zero-width limit of the LHCb analysis. A good fit does not by itself establish the molecular nature of the state; however, the fact that LQCD-constrained approaches based on the $\Bbar K$ interaction anticipated its mass, narrowness, and decay mode, together with the analogy with the $\Dsstar$, provides support for this interpretation. Since the same NLO amplitudes are being questioned in the charm sector by femtoscopic $D\pi$ correlations~\cite{ALICE:2024bhk,Albaladejo:2023pzq,Torres-Rincon:2023qll}, the bottom-strange spectrum provides an independent test in a different flavor sector. The $1^+$ partner, now correlated with the observed mass, provides a further test. Since the predicted $1^+$--$0^+$ splitting is close to the $\olsi{B}{}_s^*$--$\olsi{B}{}_s$ mass difference, the resulting doublet would be very similar if the observed peak were assigned to the $1^+$ state. Searches in the $\Bsbar\gamma$ spectrum proposed by LHCb~\cite{LHCb:2026knl} can directly test this prediction and, in turn, the chiral dynamics of heavy-light mesons in the bottom sector.

\begin{acknowledgments}
This work is part of the Grants PID2023-147458NB-C21 and CEX2023-001292-S funded by MICIU/AEI/10.13039/501100011033 and by ERDF/EU, as well as of the PROMETEO program Grant CIPROM/2023/59 funded by Generalitat Valenciana 10.13039/501100003359. %
M.\,A.\,acknowledges the \guillemotleft{}Ramón y Cajal\guillemotright{} program Grant RYC2022-038524-I funded by MICIU/AEI/10.13039/501100011033 and by ESF+, and the \guillemotleft{}Atracción de Talento\guillemotright{} program Grant PIE 20245AT019 funded by CSIC 10.13039/501100003339.
\end{acknowledgments}

\bibliographystyle{apsrev4-2-modified}
\bibliography{refs,references_modified-bibitems}

\EndMatter
\input{endmatter}

\input{supplement}

\end{document}

%% file: macros.tex
\newcommand{\papertitle}{The new LHCb beauty-strange state as the bottom partner of the $D_{s0}^*(2317)$:\\ %
a LQCD constrained coupled-channel chiral analysis}

\xspaceaddexceptions{]}
\newcommand{\modelB}{I\xspace}            
\newcommand{\modelC}{II\xspace}           
\newcommand{\modelBp}{I$'$\xspace}        
\newcommand{\modelCp}{II$'$\xspace}       
\newcommand{\fitX}{X\xspace}              
\newcommand{\fitY}{Y\xspace}              
\newcommand{\setA}{(a)\xspace}            
\newcommand{\setB}{(b)\xspace}

\newcommand{\yield}{\mathcal{N}_S}

\newcommand{\olsi}[1]{\,\overline{\!{#1}}} 

\newcommand{\Bsbar}{\olsi{B}{}_s^0}
\newcommand{\Bbar}{\olsi{B}}
\newcommand{\Bm}{B^-}
\newcommand{\Bz}{\olsi{B}{}^0}
\newcommand{\Kp}{K^+}
\newcommand{\Kz}{K^0}
\newcommand{\piz}{\pi^0}
\newcommand{\Bsstar}{\olsi{B}{}_{s0}^*}

\newcommand{\Dsstar}{D_{s0}^*(2317)}
\newcommand{\Dsone}{D_{s1}(2460)}
\newcommand{\mbare}{\mathring{m}}
\newcommand{\MeV}{\ensuremath{\,\mathrm{MeV}}\xspace}
\newcommand{\GeV}{\ensuremath{\,\mathrm{GeV}}\xspace}
\newcommand{\keV}{\ensuremath{\,\mathrm{keV}}\xspace}
\newcommand{\fm}{\ensuremath{\,\mathrm{fm}}\xspace}
\newcommand{\dd}{\mathrm{d}}

%% file: fig_specX.tex
\begin{figure}[t]
\centering
\includegraphics{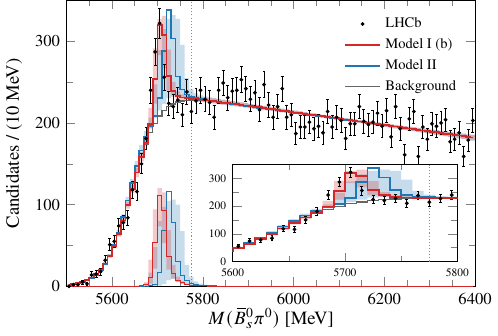}
\caption{LHCb $\Bsbar\piz$ spectrum~\cite{LHCb:2026knl} (black points) and \emph{fixed-amplitude} fits (\fitX) with Model~\modelB [set \setB, red] and Model~\modelC (blue): total (thick) and signal (thin) contributions for the central parameters, the background of the Model~\modelB fit (gray, almost indistinguishable from that of Model~\modelC), and 68\% MC bands propagating the uncertainties of the LQCD-constrained parameters. The dotted vertical line marks the $\olsi{B}K$ threshold. Model~\modelB{}\,\setA is very similar to Model~\modelB{}\,\setB and is not shown.%
\label{fig:specX}}
\end{figure}

%% file: fig_specY.tex
\begin{figure}[t]
\centering
\includegraphics{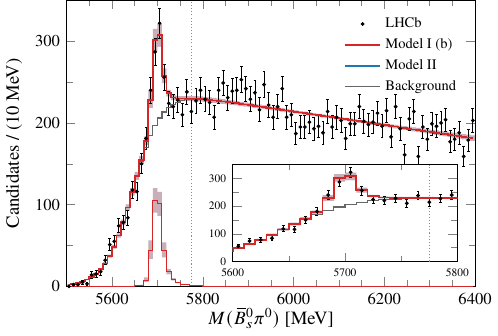}
\caption{As Fig.~\ref{fig:specX}, but for the \emph{free-amplitude} fits (\fitY), which allow $\Lambda$ (Model~\modelB) or the subtraction constant $a_b$ (Model~\modelC) to vary together with the yield and background. The two curves overlap and are indistinguishable.}
\label{fig:specY}
\end{figure}

%% file: tab_results.tex
\begin{table}[t]
\caption{Properties of the $\Bsstar$ pole for the \emph{fixed-amplitude} (\fitX) and \emph{free-amplitude} (\fitY) fits: mass, width $\Gamma(\Bsstar\to\Bsbar\piz)$, $\Bbar K$ compositeness (End Matter), and mass of the $1^+$ partner predicted with the same parameters (isospin limit). The uncertainties represent the 68\% intervals of the MC distributions.}
\label{tab:results}
\renewcommand{\arraystretch}{1.25}\small
\begin{ruledtabular}
\begin{tabular}{llcccc}
fit & model & $M_p$ [MeV] & $\Gamma$ [keV] & $X_{\Bbar K}$ [\%] & $M(1^+)$ [MeV] \\ \hline
\fitX & \modelB\ \setA & $5711.1^{+6.2}_{-6.4}$ & $41.8^{+1.4}_{-1.3}$ & $51.8^{+1.7}_{-1.5}$ & $5752.5^{+6.2}_{-6.4}$ \\
      & \modelB\ \setB & $5707.9^{+6.2}_{-6.3}$ & $34.8\pm0.6$ & $45.8^{+1.2}_{-1.1}$ & $5757.2^{+6.2}_{-6.3}$ \\
      & \modelC        & $5725.6^{+18.5}_{-22.6}$  & $96.7^{+25.9}_{-21.0}$  & $63.3^{+3.0}_{-2.6}$  & $5777.0^{+17.1}_{-21.3}$  \\ \hline
\fitY & \modelB\ \setA & $5699.6^{+1.6}_{-1.7}$ & $40.3^{+1.0}_{-0.8}$ & $48.8^{+0.6}_{-0.5}$ & $5741.0^{+1.6}_{-1.7}$ \\
      & \modelB\ \setB & $5699.6^{+1.6}_{-1.7}$ & $34.0^{+0.3}_{-0.2}$ & $44.1^{+0.6}_{-0.5}$ & $5748.8^{+1.6}_{-1.7}$ \\
      & \modelC        & $5699.6^{+1.6}_{-1.7}$ & $82.5^{+12.3}_{-10.3}$  & $59.5^{+2.0}_{-1.9}$  & $5751.7^{+1.9}_{-2.2}$  \\ \hline
\multicolumn{2}{l}{LHCb~\cite{LHCb:2026knl}} & $5698.9(1.5)(0.6)$ & $<9800$ & --- & --- \\
\end{tabular}
\end{ruledtabular}
\end{table}

%% file: endmatter.tex
\emph{LQCD input of Model~\modelB.}---The LEC $c$ and the regulator $\Lambda$ are fixed, following
Ref.~\cite{Albaladejo:2016ztm}, from the three lowest $J^P=0^+$ ($\Bbar K$) and $1^+$ ($\Bbar{}^*K$) energy levels of the
$N_f=2+1$ simulation of Ref.~\cite{Lang:2015hza} ($m_\pi=156\MeV$, $m_K=504\MeV$, lattice spacing $a^{\rm lat}=0.0907(13)\fm$,
one volume with $L=32a^{\rm lat}$). The levels are described by the isospin-symmetric, single-channel version of Model~\modelB, with the same $c$ and $\Lambda$ for $J=0$ and $1$, as required by HQSS. For each set, we use the corresponding $b\bar s$ bare masses together with the lattice meson masses and dispersion relations, given in Table~I and Eq.,(24) of Ref.,\cite{Albaladejo:2016ztm}, respectively. The finite-volume loop function is evaluated as a discrete sum over momenta $\vec q=2\pi\vec n/L$.
The energy levels $\epsilon_{J,i}$ are the solutions of $1-V(\epsilon)\widetilde G(\epsilon,L)=0$, and
\begingroup
\allowdisplaybreaks
\begin{equation}
\chi^2_{\rm LQCD}={} \sum_{\substack{i=1,2,3\\J=0,1}}\frac{\big[\epsilon_{J,i}(c,\Lambda,a^{\rm th})-\epsilon^{\rm lat}_{J,i}\big]^2}{(\Delta \epsilon_{J,i})^2}+\frac{(a^{\text{th}}-a^{\text{lat}})^2}{(\Delta a^{\text{lat}})^2},
\label{eq:chi2B}
\end{equation}
\endgroup
with the energies expressed in lattice units, $aE$, and $a^{\text{th}}$ also fitted but constrained by $a^{\text{lat}}$ as a prior~\cite{Albaladejo:2016ztm}. The uncertainties are propagated by a parametric bootstrap: the lattice levels and the central value of $a_{\rm lat}$ are resampled as independent Gaussians with their errors, and $\chi^2_{\rm LQCD}$ is minimized again~\cite{Albaladejo:2016ztm}. The resulting $(c,\Lambda)$ samples are strongly non-Gaussian and are used directly.

\emph{LQCD input of Model~\modelC.}---The LECs are taken from the fit of Ref.~\cite{Liu:2012zya} to the $S$-wave scattering lengths of $D\olsi{K}$ ($I=0,1$), $D_sK$, $D\pi$ ($I=3/2$) and $D_s\pi$ computed in LQCD on three ensembles ($m_\pi\simeq301$, 364 and $511\MeV$). The fit uses 15 data points and five free parameters: the subtraction constant and $h_{24}$, $h_4'$, $h_{35}$, $h_5'$, while $h_0$ and $h_1$ are fixed. We propagate their uncertainty with a bootstrap to the scattering lengths~\cite{Liu:2012zya}. Each set of LECs obtained in the fit is translated to the bottom sector as in Ref.~\cite{Albaladejo:2016lbb}. The 68\% intervals of the refitted parameters reproduce the errors of Ref.~\cite{Liu:2012zya}.

\begin{figure}[t]
\centering
\includegraphics{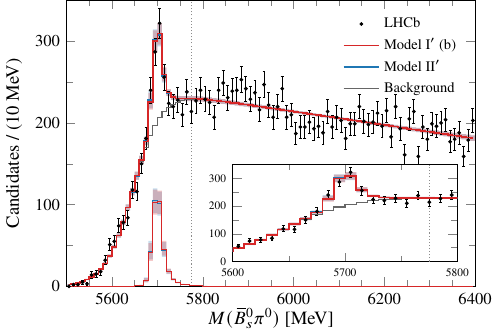}
\caption{As in Fig.~\ref{fig:specY}, the joint fits to the LHCb spectrum and the LQCD data are shown for Model~\modelBp [set \setB, red] and Model~\modelCp (blue). The bands represent the 68\% intervals obtained from the MC samples, generated by bootstrap resampling of the LQCD data and Poisson resampling of the spectrum. The predictions of the two models overlap almost completely. The results obtained with set~\setA are similar.}
\label{fig:joint}
\end{figure}
\begin{figure}[h!]
\centering
\includegraphics{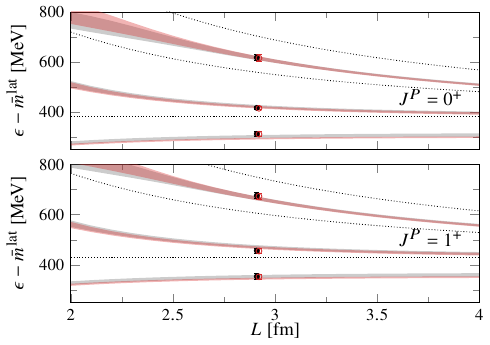}
\caption{Finite-volume $\Bbar{}^{(*)}K$ energy levels of Model~\modelB [set \setB] for $J^P=0^+$ (top) and $1^+$ (bottom) as functions of the box size $L$, measured relative to the lattice spin-averaged $\olsi{B}{}_s$ mass. The gray bands show the 68\% intervals from fits to the LQCD levels alone, while the red bands show those from the joint fit to the LQCD levels and the LHCb spectrum (\modelBp). Points show the levels of Ref.~\cite{Lang:2015hza} at $L=32a^{\rm lat}$, using the lattice spacing from the LQCD-only fit (filled) and from the joint fit (open). Dotted lines denote the non-interacting levels. The results for set \setA are similar.}
\label{fig:lqcdlevels}
\end{figure}
\begin{table*}[t]
\caption{As Table~\ref{tab:results}, for the joint fits.\label{tab:joint}}
\renewcommand{\arraystretch}{1.25}
\begin{ruledtabular}
\begin{tabular}{lcccccc}
fit & $M_p$ [MeV] & $\Gamma$ [keV] & $X_{\Bbar K}$ [\%] & $M(1^+)$ [MeV] & fitted parameters & $\Delta\chi^2_{\rm LQCD}$ \\ \hline
\modelBp\ \setA & $5700.2\pm1.6$ & $39.6^{+0.9}_{-0.6}$ & $49.3^{+0.6}_{-0.5}$ & $5741.6\pm1.6$ & $c=0.790^{+0.064}_{-0.059}$, $\Lambda=770^{+40}_{-38}\MeV$ & $2.5$ \\
\modelBp\ \setB & $5700.0\pm1.6$ & $34.0^{+0.2}_{-0.1}$ & $44.5\pm0.5$ & $5749.2\pm1.6$ & $c=0.773^{+0.043}_{-0.039}$, $\Lambda=686\pm25\MeV$ & $1.3$ \\
\modelCp        & $5699.6^{+1.6}_{-1.7}$  & $74.3^{+6.0}_{-5.8}$  & $61.0^{+1.3}_{-1.2}$  & $5752.7^{+1.6}_{-1.8}$  & $a_b=-3.453^{+0.011}_{-0.012}$ & $1.1$ \\
\end{tabular}
\end{ruledtabular}
\end{table*}

\begin{figure}[t]
\centering
\includegraphics{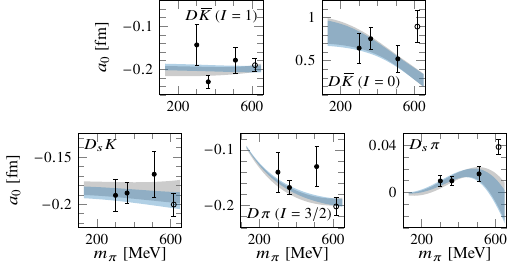}
\caption{$S$-wave scattering lengths of light pseudoscalars off charmed mesons as functions of the pion mass: LQCD data from Ref.~\cite{Liu:2012zya} (filled points: fitted; open: heaviest ensemble, not fitted) and 68\% bands of the unitarized NLO amplitudes of Model~\modelC, fitted to the lattice data alone (gray) and jointly with the LHCb spectrum (\modelCp, blue).}
\label{fig:lqcdscatt}
\end{figure}

\emph{Joint fits (variants \modelBp and \modelCp).}---In the variants \modelBp and \modelCp, the LHCb spectrum and the LQCD data are fitted simultaneously by minimizing:
\begin{equation}
D_{\rm tot}=D_{\rm LHCb}+\chi^2_{\rm LQCD},
\end{equation}
with $\chi^2_{\rm LQCD}$ of Eq.~\eqref{eq:chi2B} for \modelBp (free parameters $c$, $\Lambda$, $a^{\rm th}$, the yield $\yield$ and the background) and the $\chi^2$ of the scattering lengths of Ref.~\cite{Liu:2012zya} for \modelCp (free LECs of that fit, the yield and the background), in which the bottom subtraction constant follows from the charm one. The increase $\Delta\chi^2_{\rm LQCD}$ with respect to the fit to the LQCD data alone quantifies the tension induced by the simultaneous description of both datasets: $\Delta\chi^2_{\rm LQCD}=2.5$, $1.3$ and $1.1$ for models \modelBp{}\,\setA, \modelBp{}\,\setB, and \modelCp{}, respectively. The difference $\Delta D_{\rm tot}=\Delta D+\Delta\chi^2_{\rm LQCD}$, where $\Delta D$ is the increase of $D_{\rm LHCb}$ with respect to the reference fit, compares the joint fit with the two separate fits of the same data (the reference fit to the LHCb peak and the LQCD-only fit). We obtain $\Delta D_{\rm tot}=2.7$, $1.4$ and $1.2$. Within each model, the LHCb peak and the lattice data are thus compatible. The total $\chi^2/\mathrm{dof}$ is $1.2$ for all fits. The results and the fitted parameters are given in Tables~\ref{tab:joint} and \ref{tab:lecs}, respectively. Figures~\ref{fig:lqcdlevels} and~\ref{fig:lqcdscatt} compare the description of the lattice data in the LQCD-only and joint fits. In Model~\modelB the increase of $\chi^2_{\rm LQCD}$ comes mainly from the lowest $0^+$ level, which the joint fit places lower (the pull of the central fit changes from $-0.8$ to $-1.8$ for set \setA and from $-1.2$ to $-1.9$ for set \setB), while in Model~\modelC the bands of the two fits largely overlap. Figure~\ref{fig:joint} shows the corresponding spectra, which are indistinguishable from those of the free-amplitude fits (Fig.~\ref{fig:specY}).

\begin{table}[t]
\caption{LECs of Model~\modelC. Parameters of the charm-sector fit to the LQCD scattering lengths (with $h_0=0.01368$ and $h_1=0.42$ fixed): published best fit and errors of Ref.~\cite{Liu:2012zya}, used in Ref.~\cite{Albaladejo:2016lbb} (reproduced by our bootstrap fits), and the joint fit to the LQCD scattering lengths and the LHCb spectrum (\modelCp; central values and 16\%--84\% intervals of the MC). $a_c$ is the subtraction constant at $\mu=1\GeV$ for the charm sector, which is then translated into $a_b$ for the bottom sector as in Ref.\,\cite{Albaladejo:2016lbb}.}
\label{tab:lecs}
\renewcommand{\arraystretch}{1.3}
\begin{ruledtabular}
\begin{tabular}{lccccc}
charm & $a_c(1\GeV)$ & $h_{24}$ & $h_4'$ & $h_{35}$ & $h_5'$ \\ \hline
Ref.~\cite{Liu:2012zya} & $-1.88^{+0.07}_{-0.09}$ & $-0.10^{+0.05}_{-0.06}$ & $-0.32^{+0.35}_{-0.34}$ & $0.25^{+0.13}_{-0.13}$ & $-1.88^{+0.63}_{-0.61}$ \\
Fit \modelCp  & $-1.97^{+0.03}_{-0.03}$ & $-0.10^{+0.05}_{-0.06}$ & $-0.51^{+0.31}_{-0.30}$ & $0.16^{+0.09}_{-0.09}$ & $-1.39^{+0.42}_{-0.38}$ \\
\end{tabular}
\end{ruledtabular}
\end{table}

\emph{Monte Carlo.}---Each fit is repeated for $10000$ samples (a sample is discarded if one of its fits fails,
which happens in around one per thousand samples). In sample $k$ the parameters not fitted are
taken from the $k$-th bootstrap sample of the corresponding LQCD data, the LHCb spectrum is resampled bin by bin,
$n_b\to n_b^*\sim\mathrm{Poisson}(n_b)$, and, in the fixed-amplitude fits, the yield is drawn from the LHCb value
$343^{+44}_{-41}$. The reference fit is repeated on the same resampled spectrum. The first sample uses the central parameters, the observed spectrum and $\yield=343$. We quote the value of the first sample with the 16\%--84\% interval of the distributions, and the bands of the figures are the bin-by-bin 16\%--84\% intervals of the fitted $\mu_b$ and $S_b$.

\emph{Free-amplitude fits.}---In Model~\modelB, fitting $c$ and $\Lambda$ simultaneously is nearly degenerate, since the spectrum fixes only the combination that determines $M_p$. Therefore, we fit $\Lambda$, with $c$ taken from the bootstrap. In Model~\modelC we fit the subtraction constant $a$, whose bottom-sector value is not fitted to lattice data but obtained from the charm one by matching the loop function to a cutoff loop at the $DK$ and $\Bbar K$ thresholds~\cite{Guo:2006fu,Albaladejo:2016lbb}, with the other LECs from the bootstrap. The shift of the fitted parameter with respect to its LQCD value in the same sample measures the cost of the adjustment.

\emph{Couplings and compositeness.}---Near the pole, $T_{ij}(s)\simeq g_ig_j/(s-s_p)$; the couplings are obtained from
the residues, with a contour that does not cross the $\Bsbar\piz$ cut. The compositeness of channel $j$
is~\cite{Weinberg:1965zz,Gamermann:2009uq,Aceti:2014ala,Albaladejo:2022sux}
\begin{equation}
X_j=-g_j^2\,\frac{\dd G_j}{\dd s}\Big|_{s=s_p},\qquad Z=1-\sum_jX_j .
\label{eq:compositeness}
\end{equation}
For a bound state with energy-independent interactions $Z=0$; a nonzero $Z$ reflects the energy dependence of the
potential~\cite{Aceti:2014ala,Garcia-Recio:2015jsa}, which in Model~\modelB arises from the bare $b\bar{s}$ state. The
$X_j$ are complex because the $\Bsbar\piz$ channel is open, but their imaginary parts are negligible, and we quote the real
parts. In the isospin limit, $X_{\Bm\Kp}+X_{\Bz\Kz}$ reproduces the $\Bbar K$ probability of Ref.~\cite{Albaladejo:2016ztm}.

\emph{$1^+$ partner.}---The $1^+$ state is computed in the isospin limit, with the $\Bbar{}^*K$ channel and the same
parameters of each sample: $c$, $\Lambda$ and $\mbare_{1^+}$ in Model~\modelB~\cite{Albaladejo:2016ztm}, and the same LECs
and subtraction constant in Model~\modelC~\cite{Albaladejo:2016lbb}.

%% file: supplement.tex
\onecolumngrid
\clearpage
\setcounter{figure}{0}
\setcounter{section}{0}
\setcounter{equation}{0}
\setcounter{table}{0}
\setcounter{secnumdepth}{1}
\makeatletter
\renewcommand{\thefigure}{S\@arabic\c@figure}
\renewcommand{\thetable}{S\@arabic\c@table}
\renewcommand{\theequation}{S\@arabic\c@equation}
\renewcommand{\thesection}{S\@arabic\c@section}
\makeatother
\titleformat{\section}{\centering\small\bfseries\MakeUppercase}{\thesection.}{5pt}{}[]
\titlespacing*{\section}{0pt}{2ex}{1ex}

\begin{center}
{\bfseries\normalsize SUPPLEMENTAL MATERIAL}
\end{center}

\section{Channels and flavor coefficients}
\label{supp:flavor}
The channels are $1=\Bsbar\piz$, $2=\Bm\Kp$, $3=\Bz\Kz$ and $4=\Bsbar\eta$ (Model~\modelB: channels 1--3).\footnote{The $\Bbar K$--$\Bsbar\pi$ coupled
channels were also studied, with $I=1$, in connection with the $X(5568)$ structure~\cite{Albaladejo:2016eps,Lu:2016kxm,Guo:2016nhb}.} The
masses are those of the PDG~\cite{PDG2026}, and the thresholds are at $5501.9$, $5773.1$, $5777.3$ and $5914.8\MeV$,
respectively. At LO the neutral light-meson mass matrix in the $(\pi_3,\eta_8)$ basis has
an off-diagonal element $B(m_u-m_d)/\sqrt3$, and the physical states are
\begin{equation}
\piz=\cos\varepsilon\,\pi_3+\sin\varepsilon\,\eta_8,\qquad \eta=-\sin\varepsilon\,\pi_3+\cos\varepsilon\,\eta_8,
\qquad \tan2\varepsilon=\frac{\sqrt3\,(1-r)}{(1+r)(S-1)},
\end{equation}
with $r=m_u/m_d=0.465(24)$, $S=m_s/m_{ud}=27.227(81)$~\cite{FLAG:2024}, and $m_{ud}$ the isospin-averaged up- and down-quark mass, giving $\varepsilon=0.01206$. Only the angle
is taken from the LO relation; the masses are the physical ones. With the Goldstone-boson matrix
$\phi=\sum_k\lambda_k\phi_k$ and $\lambda_{\piz}=\cos\varepsilon\,\lambda_{\pi_3}+\sin\varepsilon\,\lambda_{\eta_8}$,
$\lambda_\eta=-\sin\varepsilon\,\lambda_{\pi_3}+\cos\varepsilon\,\lambda_{\eta_8}$, the WT coefficient for
$P_{a_i}\phi_i\to P_{a_j}\phi_j$ ($P_a=\Bm,\Bz,\Bsbar$) is $C_{ij}=[\lambda_j^T,\lambda_i]_{a_ia_j}$:
\begin{equation}
C=\begin{pmatrix}
0 & \tfrac{c_\varepsilon}{\sqrt2}+\sqrt{\tfrac32}s_\varepsilon & -\tfrac{c_\varepsilon}{\sqrt2}+\sqrt{\tfrac32}s_\varepsilon & 0\\[3pt]
\cdot & -1 & -1 & \sqrt{\tfrac32}c_\varepsilon-\tfrac{s_\varepsilon}{\sqrt2}\\[3pt]
\cdot & \cdot & -1 & \sqrt{\tfrac32}c_\varepsilon+\tfrac{s_\varepsilon}{\sqrt2}\\[3pt]
\cdot & \cdot & \cdot & 0
\end{pmatrix},
\end{equation}
with $c_\varepsilon=\cos\varepsilon$, $s_\varepsilon=\sin\varepsilon$, and $C_{ji}=C_{ij}$. In the isospin limit the $I=0$
combination $(\Bm\Kp+\Bz\Kz)/\sqrt2$ has $C=-2$ and decouples from $\Bsbar\piz$. In Eq.~\eqref{eq:V},
$u_{ij}=m_{H,i}^2+m_{L,j}^2-2E_{H,i}E_{L,j}$, with the on-shell c.m.\ energies of the mesons,
$E_{H,i}=(s+m_{H,i}^2-m_{L,i}^2)/(2\sqrt s)$ and $E_{L,i}=(s+m_{L,i}^2-m_{H,i}^2)/(2\sqrt s)$. The bare $b\bar{s}$ state couples
to $P_{a_i}\phi_i$ with $F_i=(\lambda_i)_{a_i3}$: $F_{\Bsbar\piz}=-\tfrac{2}{\sqrt6}\sin\varepsilon$, $F_{\Bm\Kp}=F_{\Bz\Kz}=1$,
normalized so that the second term of Eq.~\eqref{eq:V} reproduces, in the isospin limit, the $I=0$ potential of
Ref.~\cite{Albaladejo:2016ztm} [written there with $(s-m_{H,i}^2+m_{L,i}^2)/(2\sqrt s)=E_{L,i}$]. The NLO coefficients of Eq.~\eqref{eq:V} follow from the $\mathcal O(p^2)$
Lagrangian of Refs.~\cite{Guo:2008gp,Liu:2012zya} with the same $\lambda_k$ and $\chi=\mathrm{diag}(x_u,x_d,x_s)$, $x_q=2Bm_q$:
\begin{gather}
C^0_{ij}=\frac{\delta_{a_ia_j}}2\,\mathrm{Tr}\big(\{\lambda_j^T,\lambda_i\}\chi\big),\qquad
C^1_{ij}=-\frac14\Big(\big\{\lambda_j^T,\{\lambda_i,\chi\}\big\}+\big\{\lambda_i,\{\lambda_j^T,\chi\}\big\}\Big)_{a_ia_j},\\
C^{24}_{ij}=\frac{\delta_{a_ia_j}}2\,\mathrm{Tr}\big(\{\lambda_j^T,\lambda_i\}\big),\qquad
C^{35}_{ij}=\big\{\lambda_j^T,\lambda_i\big\}_{a_ia_j}.
\end{gather}
In the isospin limit, these expressions reproduce the coefficients of Ref.~\cite{Liu:2012zya}. The direct isospin-breaking
part is obtained with $\delta\chi=\mathrm{diag}(\delta x,-\delta x,0)$, $\delta x=m_\pi^2(r-1)/(r+1)$.

\section{NLO potential and LECs}
\label{supp:nlo}
In Eq.~\eqref{eq:V}, for $P(p_1)\phi(p_2)\to P(p_3)\phi(p_4)$~\cite{Liu:2012zya},
\begin{equation}
H_{24}=2h_{24}\,p_2\!\cdot\!p_4+\frac{h_4'}{\bar M^2}\big(p_1\!\cdot\!p_2\,p_3\!\cdot\!p_4+p_1\!\cdot\!p_4\,p_2\!\cdot\!p_3-2\bar M^2\,p_2\!\cdot\!p_4\big),\quad
H_{35}=h_{35}\,p_2\!\cdot\!p_4+\frac{h_5'}{\bar M^2}\big(p_1\!\cdot\!p_2\,p_3\!\cdot\!p_4+p_1\!\cdot\!p_4\,p_2\!\cdot\!p_3-2\bar M^2\,p_2\!\cdot\!p_4\big),
\end{equation}
projected onto the $S$ wave, with $h_{24}=h_2+h_4'$, $h_{35}=h_3+2h_5'$ and $\bar M=(m_D+m_{D_s})/2$, where $m_D$ and $m_{D_s}$ are the physical $D$ and $D_s$ meson masses. Following Ref.~\cite{Albaladejo:2016lbb}, $h_{0\text{--}3}$ scale with the heavy-meson mass and $h'_{4,5}$ with its inverse. The charm-sector values are given in Table~\ref{tab:lecs}.

\section{Loop functions}
\label{supp:loops}
With $\omega_i=\sqrt{m_i^2+\vec q^{\,2}}$ and $p$ the c.m.\ momentum, the Gaussian-regularized loop of
Model~\modelB~\cite{Albaladejo:2016ztm} is
\begin{equation}
G_\Lambda(s)=\int\frac{\dd^3q}{(2\pi)^3}\,\frac{\omega_1+\omega_2}{2\omega_1\omega_2}\,
\frac{e^{-2(\vec q^{\,2}-p^2)/\Lambda^2}}{s-(\omega_1+\omega_2)^2+i\epsilon},
\end{equation}
and the dimensionally regularized loop of Model~\modelC~\cite{Oller:2000fj} is
\begin{multline}
16\pi^2G(s)=a(\mu)+\ln\frac{m_1m_2}{\mu^2}+\frac{\Delta}{2s}\ln\frac{m_2^2}{m_1^2}
+\frac{p}{\sqrt s}\Big[\ln\big(s-\Delta+2\sqrt s\,p\big)+\ln\big(s+\Delta+2\sqrt s\,p\big)\\
-\ln\big(-s+\Delta+2\sqrt s\,p\big)-\ln\big(-s-\Delta+2\sqrt s\,p\big)\Big],
\end{multline}
with $\Delta=m_2^2-m_1^2$.
Both are normalized so that $\mathrm{Im}\,G(s+i0)=-p/(8\pi\sqrt s)$. On the second sheet,
$G^{\rm II}(s)=G(s)+i\,p/(4\pi\sqrt s)$, with $\mathrm{Im}\,p\ge0$. The $\Bsstar$ pole lies on the sheet in which only
the $\Bsbar\piz$ loop is continued.

\section{Production}
\label{supp:production}
In Eq.~\eqref{eq:prod} we use $\alpha_{\Bsbar\piz}=\alpha_{\Bsbar\eta}=0$ and $\alpha_{\Bm\Kp}=\alpha_{\Bz\Kz}$. With a fixed
amplitude, a free direct term $\alpha_{\Bsbar\piz}$ (or $\alpha_{\Bsbar\eta}$) interferes destructively with the resonant
term and turns the fixed yield into a smooth continuum absorbed by the background, hiding a displaced pole; with a
fitted amplitude these terms change neither the fit quality nor the pole appreciably. A difference
$\alpha_{\Bm\Kp}\neq\alpha_{\Bz\Kz}$ only feeds an $I=1$ continuum and is degenerate with the yield. In summary, the
spectrum determines the pole mass and the number of events in the peak, Eq.~\eqref{eq:Npeak}, but not how the
production is shared among the channels.

\section{Resolution function}
\label{supp:resolution}
\begin{figure}[h]
\centering
\includegraphics{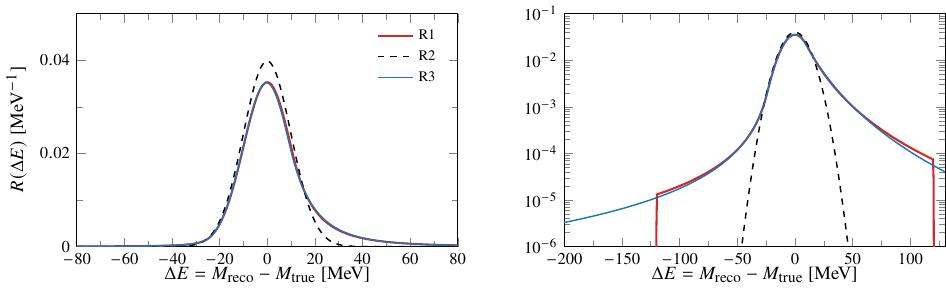}
\caption{Resolution kernels (unit area): R1, the signal component of the LHCb fit to the spectrum summed over the three decay
modes~\cite{LHCb:2026knl} (red); R2, a Gaussian with the same full width at half maximum (dashed); R3, a double-sided Crystal
Ball function fitted to R1 (blue). Left: linear scale; right: logarithmic scale, showing the tails.}
\label{fig:kernels}
\end{figure}

The kernel $R$ of Eq.~\eqref{eq:R} (R1) is the signal curve of Fig.~1 of Ref.~\cite{LHCb:2026knl}, extracted from the vector graphics of the figure, normalized to unit area and centered at $M_{\rm LHCb}$. It has a full width at half maximum of $23.6\MeV$, sizeable tails, and is defined in $|E-M_{\rm LHCb}|\lesssim120\MeV$. As checks, we use a Gaussian with $\sigma=10.02\MeV$ (R2), chosen to have the same FWHM, and a double-sided Crystal Ball function~\cite{Skwarnicki:1986xj} fitted to R1 (R3) (Fig.~\ref{fig:kernels}). Repeating all the fits with R2 and R3 changes the pole masses and the $1^+$ masses by at most $0.5\,\MeV$, the widths by at most $0.4\keV$, and the $\Bbar K$ compositeness by at most $0.001$. The yield is practically unchanged with R3. With R2 it decreases by up to $37$ events in all the fits, including the reference one, because a Gaussian kernel without tails assigns a different fraction of the signal to the peak.

\section{Expected counts and normalization}
\label{supp:counts}
The expected number of candidates in bin $b$ ($b=1,\dots,90$, bins of $10\MeV$ in $[5500,6400]\MeV$) is
\begin{equation}
\mu_b=S_b+B_b,\qquad S_b=\yield\,\frac{s_b}{\sum_{b'=1}^{90}s_{b'}},\qquad
s_b=\int_b\dd\sqrt s\int\dd E\,R(\sqrt s-E)\,\frac{\dd N}{\dd E},
\label{eq:Sb}
\end{equation}
where $\dd N/\dd E\propto(p_1/E)\,|\mathcal A(E)|^2$ is the theoretical distribution of Eq.~\eqref{eq:prod}, with $E$ the
invariant mass, and $R$ is normalized to unit area. The overall normalization of $\dd N/\dd E$ cancels in $S_b$, and
$\yield=\sum_bS_b$ is the number of signal candidates in the fit window. Since the pole is narrow compared with $R$,
$S_b\simeq\yield\int_b\dd\sqrt s\,R(\sqrt s-M_p)$ up to the small non-resonant continuum. For the reference fit,
$S_b=\yield\int_b\dd\sqrt s\,R(\sqrt s-M_0)$. The background is
\begin{equation}
B_b=\int_b\dd\sqrt s\,N_b\,\frac{1+b_1x}{1+e^{-(\sqrt s-m_0)/w}},\qquad x=\frac{\sqrt s-5950\MeV}{450\MeV},
\end{equation}
with the four parameters $(N_b,b_1,m_0,w)$ always free~\cite{LHCb:2026knl}.

\section{Scattering lengths and couplings}
\label{supp:observables}
Table~\ref{tab:observables} gives the couplings of the $\Bsstar$ pole, the compositeness of the two $\Bbar K$ channels,
$Z$, and the scattering lengths, defined by $T_{jj}=-8\pi\sqrt s\,a_j$ at the threshold of channel $j$. The $\Bm\Kp$
scattering length has a sizeable imaginary part: its $I=1$ component couples to the open $\Bsbar\piz$ channel without
isospin violation. The isoscalar scattering length $a(I=0)$ is computed in the isospin limit, where it is real.
\begin{table}[h]
\caption{Couplings, compositeness and scattering lengths for all fits (R1). Central values and 16\%--84\% intervals
of the MC. For Model~\modelC, $X_{\Bsbar\eta}=21.7^{+0.7}_{-1.4}\%$ (\fitX), $21.6\pm0.9\%$ (\fitY) and $21.1\pm0.8\%$ (\modelCp),
respectively.}
\label{tab:observables}
\renewcommand{\arraystretch}{1.3}\footnotesize
\begin{ruledtabular}
\begin{tabular}{llcccccccc}
fit & model & $|g_{\Bsbar\piz}|$ [GeV] & $|g_{\Bm\Kp}|$ [GeV] & $|g_{\Bz\Kz}|$ [GeV] & $X_{\Bm\Kp}$ [\%] & $X_{\Bz\Kz}$ [\%] & $Z$ [\%] & $a_{\Bm\Kp}$ [fm] & $a(I=0)$ [fm] \\ \hline
\fitX & \modelB\ \setA & $0.334\pm0.003$ & $22.45^{+0.67}_{-0.63}$ & $22.38^{+0.66}_{-0.62}$ & $26.8^{+0.9}_{-0.8}$ & $25.0\pm0.7$ & $48.2^{+1.5}_{-1.7}$ & $-0.537^{+0.031}_{-0.036}+i\,0.094\pm0.005$ & $-0.90^{+0.04}_{-0.05}$ \\
 & \modelB\ \setB & $0.306\pm0.000$ & $22.81^{+0.59}_{-0.58}$ & $22.77^{+0.59}_{-0.57}$ & $23.7^{+0.7}_{-0.6}$ & $22.1\pm0.5$ & $54.2^{+1.1}_{-1.2}$ & $-0.536^{+0.030}_{-0.035}+i\,0.089\pm0.004$ & $-0.88^{+0.04}_{-0.05}$ \\
 & \modelC & $0.498^{+0.051}_{-0.045}$ & $16.05^{+1.36}_{-1.58}$ & $15.87^{+1.36}_{-1.54}$ & $32.7^{+1.9}_{-1.5}$ & $30.6\pm1.1$ & $15.0^{+2.7}_{-2.4}$ & $-0.431^{+0.087}_{-0.148}+i\,0.141^{+0.022}_{-0.030}$ & $-0.74^{+0.13}_{-0.21}$ \\ \hline
\fitY & \modelB\ \setA & $0.334^{+0.004}_{-0.003}$ & $22.78^{+0.53}_{-0.45}$ & $22.68^{+0.54}_{-0.46}$ & $25.2\pm0.3$ & $23.6^{+0.3}_{-0.2}$ & $51.2^{+0.5}_{-0.6}$ & $-0.476^{+0.016}_{-0.018}+i\,0.104\pm0.004$ & $-0.81^{+0.02}_{-0.03}$ \\
 & \modelB\ \setB & $0.307^{+0.001}_{-0.000}$ & $22.95^{+0.51}_{-0.45}$ & $22.89^{+0.52}_{-0.46}$ & $22.8\pm0.3$ & $21.3^{+0.3}_{-0.2}$ & $55.9^{+0.5}_{-0.6}$ & $-0.488^{+0.017}_{-0.018}+i\,0.096\pm0.003$ & $-0.82^{+0.02}_{-0.03}$ \\
 & \modelC & $0.478^{+0.034}_{-0.031}$ & $17.39^{+0.28}_{-0.27}$ & $17.17^{+0.30}_{-0.29}$ & $30.5^{+1.0}_{-0.9}$ & $28.9\pm1.0$ & $19.0\pm1.2$ & $-0.348^{+0.017}_{-0.027}+i\,0.172^{+0.018}_{-0.028}$ & $-0.59\pm0.01$ \\ \hline
\fitY & \modelBp\ \setA & $0.331^{+0.004}_{-0.002}$ & $23.29^{+0.65}_{-0.56}$ & $23.20^{+0.66}_{-0.57}$ & $25.4\pm0.3$ & $23.8^{+0.3}_{-0.2}$ & $50.7^{+0.5}_{-0.6}$ & $-0.495^{+0.018}_{-0.020}+i\,0.100\pm0.004$ & $-0.84^{+0.02}_{-0.03}$ \\
 & \modelBp\ \setB & $0.306^{+0.001}_{-0.000}$ & $23.34^{+0.55}_{-0.50}$ & $23.29^{+0.56}_{-0.50}$ & $23.0\pm0.3$ & $21.5\pm0.2$ & $55.5\pm0.5$ & $-0.503^{+0.017}_{-0.018}+i\,0.093\pm0.003$ & $-0.84\pm0.02$ \\
 & \modelCp & $0.453\pm0.018$ & $17.59^{+0.19}_{-0.18}$ & $17.39^{+0.20}_{-0.19}$ & $31.3\pm0.6$ & $29.7\pm0.6$ & $17.9\pm0.6$ & $-0.332^{+0.008}_{-0.012}+i\,0.155^{+0.021}_{-0.023}$ & $-0.60\pm0.01$ \\
\end{tabular}
\end{ruledtabular}
\end{table}

\section{The higher pole of Model~\modelB}
\label{supp:resonance}
The model with the WT interaction and the exchange of a bare state also contains a broad resonance above the $\Bbar K$
thresholds, at around $6.2$--$6.3\GeV$ with a width of about $80\MeV$~\cite{Albaladejo:2016ztm}, which leaves no visible
trace in the $\Bsbar\piz$ spectrum. This is the mirror image of what happens for the bound state: below the $\Bbar K$
thresholds, $\Bsbar\piz$ is the only open channel and receives the whole yield, independently of the smallness of its
isospin-violating coupling, whereas above them the resonance decays almost exclusively into the isospin-allowed
$\Bbar K$ channels, with a $\Bsbar\piz$ branching ratio of order $10^{-6}$. The expected number of $\Bsbar\piz$ events
from this state in the LHCb sample is negligible, as we have explicitly checked.